\documentclass[manuscript]{aastex}
\usepackage{color}
\begin{document}
\newcommand{\changeP}[1]{{\textcolor{blue}{#1}}}

\received{}
\accepted{}
\journalid{}{}
\articleid{}{}
\paperid{}
\cpright{AAS}{}
\ccc{}

\slugcomment{Submitted \today\ to the {\it Astrophysical Journal}}

\shorttitle{Rotational quenching of  due to He}
\shortauthors{Yang et al.}


\title{QUENCHING OF HIGHLY-ROTATIONALLY-EXCITED HCL IN COLLISIONS WITH HE } 

\author{Benhui Yang and P. C. Stancil}
   \affil{Department of Physics and Astronomy and the Center for 
    Simulational Physics,\\  The University of Georgia,
    Athens, GA 30602, USA}
    \email{yang@physast.uga.edu, stancil@physast.uga.edu}

\begin{abstract}
We report rotational quenching cross sections and rate coefficients of HCl 
due to collisions with He. 
The close-coupling method and the coupled-states approximation are applied in 
quantum-mechanical scattering calculations of 
state-to-state cross sections for HCl with initial rotational 
levels up to $j=20$ for kinetic energies from 10$^{-5}$ to 15000 cm$^{-1}$. 
State-to-state rate coefficients for temperature between 0.1 and 3000 K are also presented. 
Comparison of the present rate 
coefficients with previous results reported in the literature 
for lowly-excited rotational levels shows reasonable agreement. 
Small differences 
are attributed to the differences in the interaction potential energy surfaces.
The uncertainty in the computed cross sections
and rate coefficients is estimated by varying the potential well depth.
Applications of current results to astrophysical systems are also briefly discussed.

\end{abstract}


\section{INTRODUCTION}

Molecular collisional excitation rate coefficients are important as they contribute to level
excitation (and quenching or deexcitation) in
competition with radiative processes. In cold environments such as the ISM,
the dominant collision partners are often H$_2$ and He,
except in photodissociation regions (PDRs) and diffuse gas where collisions
with electrons and H can become important.
Precise laboratory data including collisional (de)-excitation rate coefficients
are required, for a range of temperatures,
to interpret the complicated interstellar spectra of molecular gas not in 
local thermodynamic equilibrium (LTE).
Because of the complexity and difficulty in direct measurements, only limited 
state-to-state collision rate coefficients have been
obtained for systems of astrophysical interest \citep[see, for example,][]{car04}.
Therefore  modeling astrophysical systems rely heavily on theoretical estimates 
\citep{flo07,dub09,liq09,dan11,klo12,fau12,liq12,wie13,rou13,yang13,dub13}.

Here we consider the hydrogen halide HCl, which
 has been detected in the atmospheres of some planets, as well as in
interstellar clouds \citep{bla85,sal96,bah01}. 
It is an important tracer of chlorine and can be used to constrain 
the chlorine elemental abundance and isotope ratios.
 \citet{zmu95} observed it toward the Sagittarious B2 complex and modeled its non-LTE spectrum.
Further, HCl was recently detected in the circumstellar envelope 
of IRC+10216 by {\it Herschel} \citep{cer10}. It has also been surveyed throughout the interstellar
medium (ISM) via 
its rotational line emission by \citet{pen10}. It appears that HCl is a particularly good tracer 
of molecular cloud cores of the highest density; however, the density estimates are limited by the 
uncertainty in collision rates.

The HCl-He collision system has been investigated theoretically
and experimentally in a variety of studies \citep{gre75,col76,hel80,neu94,lan12,aji13}. 
The first reliable potential energy surface (PES) for the HCl-He complex  was presented by 
\citet{wil92} and used to theoretically fit experimental helium pressure
broadening cross sections of DCl at very low temperature and at 300 K. 
\citet{neu94} computed HCl-He rate coefficients for
temperatures between 10 K and 300 K on the potential by \citet{wil92},
but their calculations were limited to $j<8$, where $j$ is the rotational quantum number. 
In their calculations,  collision cross sections were computed using 
the MOLSCAT scattering code  \citep{molscat}
within the essentially exact close-coupling (CC) formalism.  
For total energies above 1200 K, the coupled-states (CS) approximation was used.
Two newer PESs for HCl-He have been reported \citep{mur04,faj06}.
The two-dimensional (rigid monomer) PES of \citet{mur04} was developed
from ab initio calculations using symmetry-adapted perturbation theory (SAPT) with  
an ($spdfg$) basis set plus midbond functions. This SAPT potential is in good agreement with
the semiempirical PES of \citet{wil92} predicting that the global minimum is on the Cl side. 
The accuracy of the SAPT potential was tested by performing rovibrational bound states calculations
of the HCl-He complex, which
show that the calculated transition frequencies were in excellent agreement with the measurements
of \citet{lov90}. \citet{mur04} also pointed out that the dissociation energy predicted by the 
SAPT potential for the complex is probably
more accurate than the experiment.
\citet{faj06} presented another two-dimensional HCl-He potential 
using the coupled-cluster with single, double, and 
triple excitations (CCSDT) method. This potential is very  similar to the SAPT PES, but gives better
agreement with the experimental dissociation energy and rotational energy 
levels of the HCl-He complex. 
Very recently three-dimensional and two-dimensional PESs were presented by \citet{lan12}
and by \citet{aji13}, which used the supermolecular approach based on the CCSD(T) method, but at 
different levels of theory. These new potentials were also used in CC 
calculations of pure rotational excitation of HCl by He by these authors.
 Cross sections for transitions among
the first 11  rotational states of HCl were calculated for total energy of up to 3000 cm$^{-1}$ using
MOLSCAT with the log-derivative propagator of \citet{man86}.
Rate coefficients were presented from 5 to 300 K.

In this work, we performed explicit quantum scattering calculations of
rotational quenching of HCl by He to higher levels of rotational excitation using the
SAPT potential of \citet{mur04}.  HCl-He rate coefficients are presented for a large range of 
temperature (0.1-3000 K) which will aid
in  modeling rotational spectra of HCl in various astrophysical and atmospheric environments.
We discuss the computational method in Section 2 and the results in Section 3. In Section 4,
an estimate of the uncertainty in the cross sections and rate coefficients is presented, while 
Section 5 briefly discusses application of the current results to astrophysics.

\section{COMPUTATIONAL METHOD}

The theory developed by \citet{art63} for atom-diatom scattering is adopted.
HCl was treated as a rigid-rotor with  bond length equal to the equilibrium distance
$r_e=1.275$~\AA.
The calculations presented here were performed by applying the CC method and the CS approximation
\citep[see, for example,][]{flo07}.
The HCl-He interaction potential is expressed by $V(R,\theta)$, 
where $R$ is the distance from the HCl center of mass to the He atom,
and  $\theta$ is the angle between $\vec{R}$ and the HCl molecular axis,
with $\theta = 180 ^{\circ}$ defined for the collinear arrangement He-H-Cl.
 The potential $V(R, \theta)$ was expanded in the form
\begin{equation}
V(R,\theta)=\sum_{\lambda=0}^{\lambda_{\textrm{max}}}v_{\lambda}(R)P_{\lambda}(\textrm{cos} \theta),
\end{equation}
where $P_{\lambda}$ are  Legendre polynomials and $v_{\lambda}(R)$ expansion coefficients
of the potential. 

For a transition from an initial rotational state  $j$ to a final rotational state  $j'$,
the integral cross section can be expressed in terms of
the scattering matrix $S$, within the CC formalism by
\begin{equation}
\sigma_{j\rightarrow j'}(E_{j})
=\frac{\pi}{(2j+1)k_{j}^2}\sum_{J=0}(2J+1)\sum_{l=|J-j|}^{J+j}
\sum_{l'=|J-j'|}^{J+j'}|\delta_{jj'}\delta_{ll'}
-S_{jj'll'}^J(E_j)|^2,
\end{equation}
where the total angular momentum $\bf{\vec{J}=\vec{l}+\vec{j}}$, is composed of the rotational angular
momentum $\bf{\vec{j}}$ of the HCl molecule and the orbital angular momentum $\bf{\vec{l}}$ of 
the collision complex.  
 $k_{j}=\sqrt{2\mu E_{j}}/\hbar$  is the wave vector    
for the incoming channel,  $E_j$ the center-of-mass kinetic energy for the 
incoming channel corresponding to the initial rotational state $j$ of HCl, and $\mu$ the collision
system reduced mass.

All reported scattering calculations were performed using
the quantum-mechanical scattering code MOLSCAT \citep{molscat}.
The propagation was carried out from an intermolecular separation $R=1.0$ to 
$R=100$  \AA.  To  ensure the accuracy of the 
state-to-state rate constants for temperatures from $10^{-4}$ to 3000 K, 
kinetic energies between
 $10^{-5}$ cm$^{-1}$ and 15,000 cm$^{-1}$ were used in our state-to-state cross section calculations.
The angular dependence of the interaction potential was expanded in Legendre polynomials 
shown in Eq.~(1) with $\lambda_{\textrm{max}}$=22.
24 points in $\theta$ from Gauss-Legendre quadrature were used to project
out the potential expansion coefficients.
Sufficient number of partial waves necessary for convergence of the cross sections were used; 
in the higher collision energy region the maximum value of $J$ employed was 360.
The CS approximation was adopted for collision energies greater than 2000 cm$^{-1}$,
 while the CC method was used for all lower energies, 
 the agreement between the CS and CC calculations is within 5\%.

\section{RESULTS AND DISCUSSION}

\subsection{State-to-state and total deexcitation cross sections}

\begin{figure}
\epsscale{0.9}
\plotone{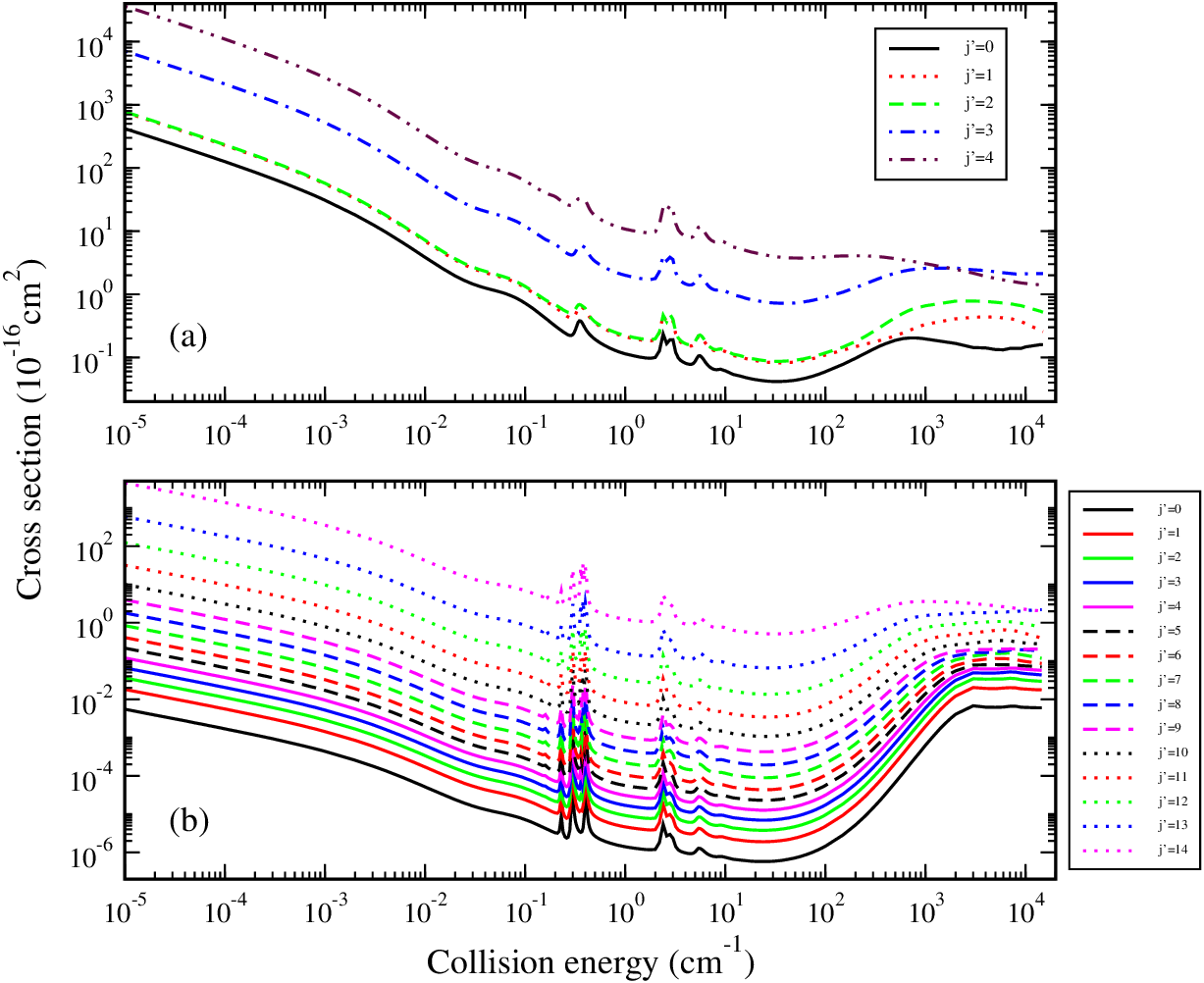}
\caption{
State-to-state rotational deexcitation cross sections of HCl due to
 collisions with He as a function 
of kinetic energy. Initial rotational state (a) $j=5$ and (b) $j=15$. 
}
\label{fig1}
\end{figure}

State-to-state quenching cross sections were computed for initial HCl rotational
levels of  $j=1, \  2, \ \cdots, \ 20$. Rotational energy levels are
 presented in Table 1 for the basis set adopted in the current calculation,
which were obtained using rotational constant $B_0$=10.5933 cm$^{-1}$ \citep{iri07} and
centrifugal distortion constant $D_0$=0.00053 cm$^{-1}$ \citep{ran65}.
As examples, the state-to-state quenching cross sections from initial levels  
$j$=5 and 15 are presented in Fig.~\ref{fig1}(a) and \ref{fig1}(b),
respectively.\footnote{All state-to-state 
deexcitation cross sections and rate coefficients 
are available on the UGA Molecular Opacity Project website
(www.physast.uga.edu/ugamop/). The rate coefficients are also available in
the BASECOL \citep{dub06} 
and the Leiden Atomic and Molecular Database (LAMDA) \citep{sch05} formats.} 
Cross sections using the CS approximation begin at 2000 cm$^{-1}$ and are seen to be 
in excellent agreement with those obtained with the CC method. 

The cross sections display resonances 
in the intermediate energy region from $\sim$0.1 cm$^{-1}$ to $\sim$10 cm$^{-1}$ due to 
quasibound levels supported by the attractive part of the interaction potential. 
The  $|\Delta j|=|j'-j| = 1$  transition dominants the 
quenching for all $j$ (shown here for $j=5$ and $j=15$), 
with the cross sections generally increasing
with increasing $j'$ with that for $j'=0$ being the smallest.

\begin{figure}
\epsscale{0.9}
\plotone{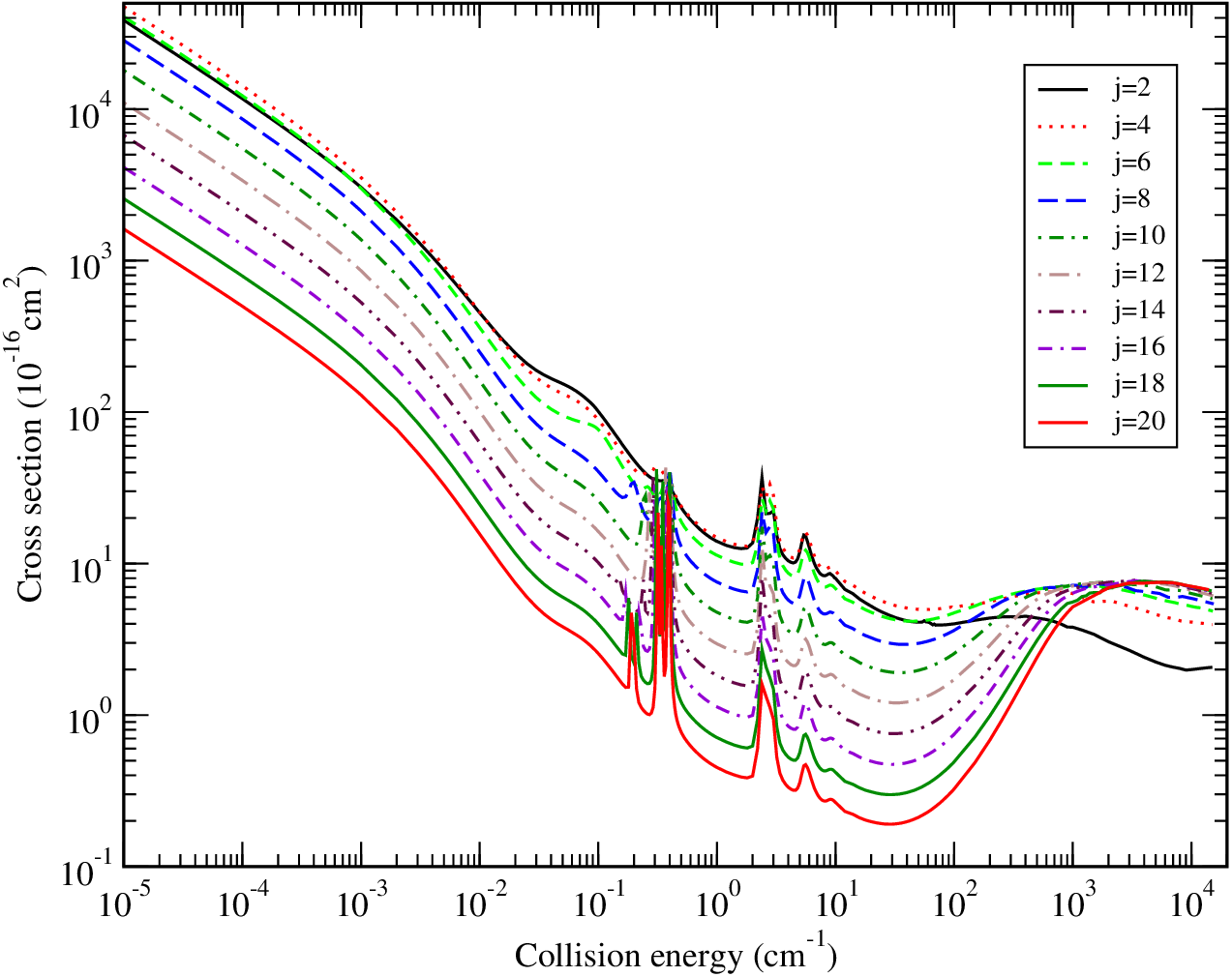}
\caption{
Total deexcitation cross sections from initial states $j$=2, 4, 6, 8, 10, 12, 14, 16, 18, 
and 20 of HCl due to collisions with He as a function of kinetic energy.
}
\label{fig2}
\end{figure}

The total deexcitation cross section from an initial state $j$ can be obtained 
by summing over all final states $j'$. 
Fig.~\ref{fig2} displays the total deexcitation cross section for quenching from
selected initial levels $j$=2, 4, 6, $\cdots$, 18, and 20.  
Generally, the total quenching cross sections 
have similar behavior, but differences result for small $j$ at high energy due to a
limited number of final exit channels. 
Each of the cross sections 
exhibits the behavior predicted by Wigner (1948) threshold laws
 at ultra-low collision energies below $\sim$10$^{-4}$ cm$^{-1}$, 
where only $s$-wave scattering
contributes and the cross sections vary inversely with the relative velocity. 
Except for $j$=2 the total deexcitation cross sections decrease 
to a global minimum near 500 cm$^{-1}$.

\subsection{State-to-state deexcitation rate coefficients}

\begin{figure}
\epsscale{0.9}
\plotone{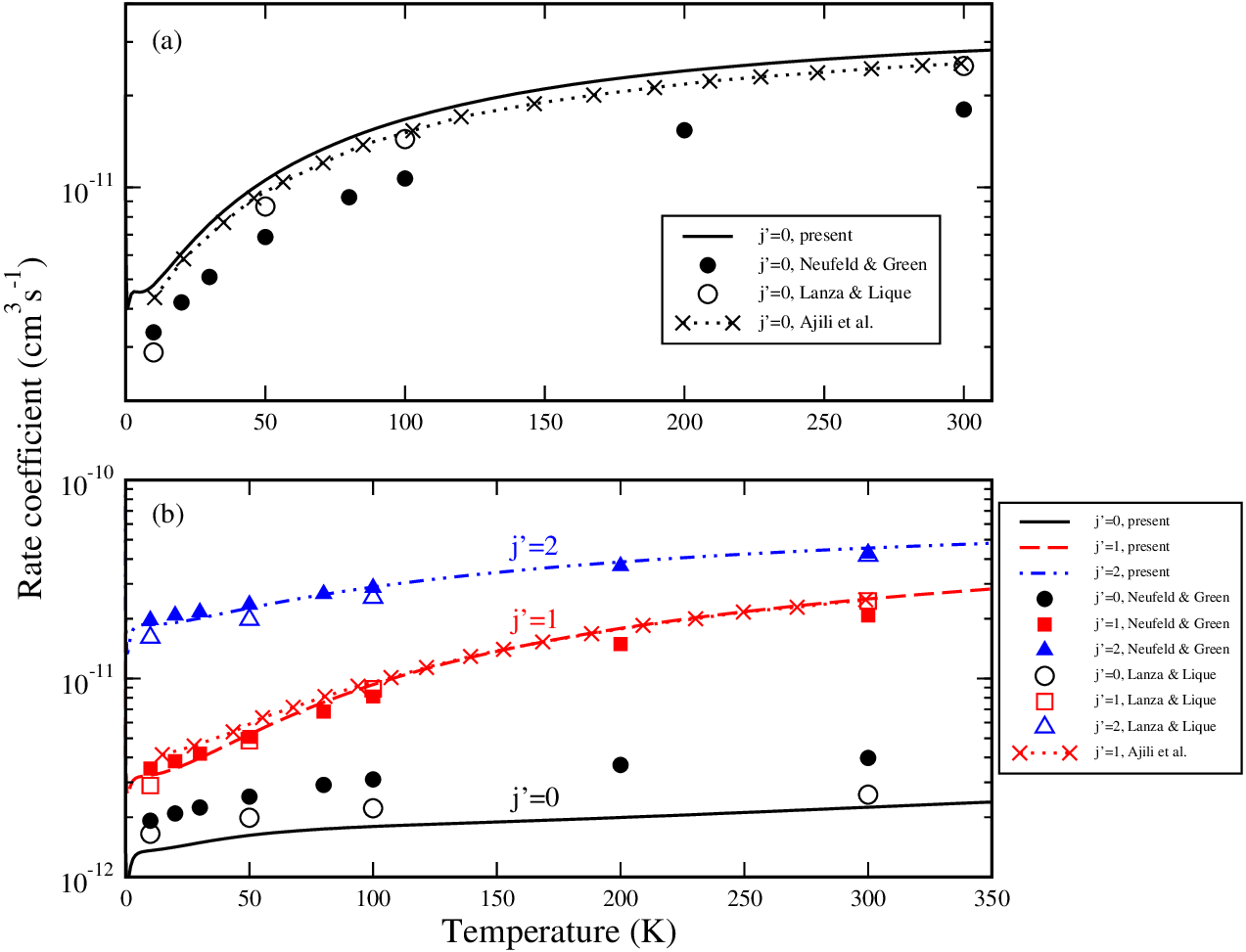}
\caption{ 
State-to-state rotational deexcitation rate coefficients from initial states 
$j$=1 and 3 of HCl due to collisions with He. (a) $j=1$, (b) $j=3$.
Lines:  present results; filled symbols: \citet{neu94}; open symbols: \citet{lan12};
dotted line with cross symbols: \citet{aji13}.
}
\label{fig3}
\end{figure}

\begin{figure}
\epsscale{0.9}
\plotone{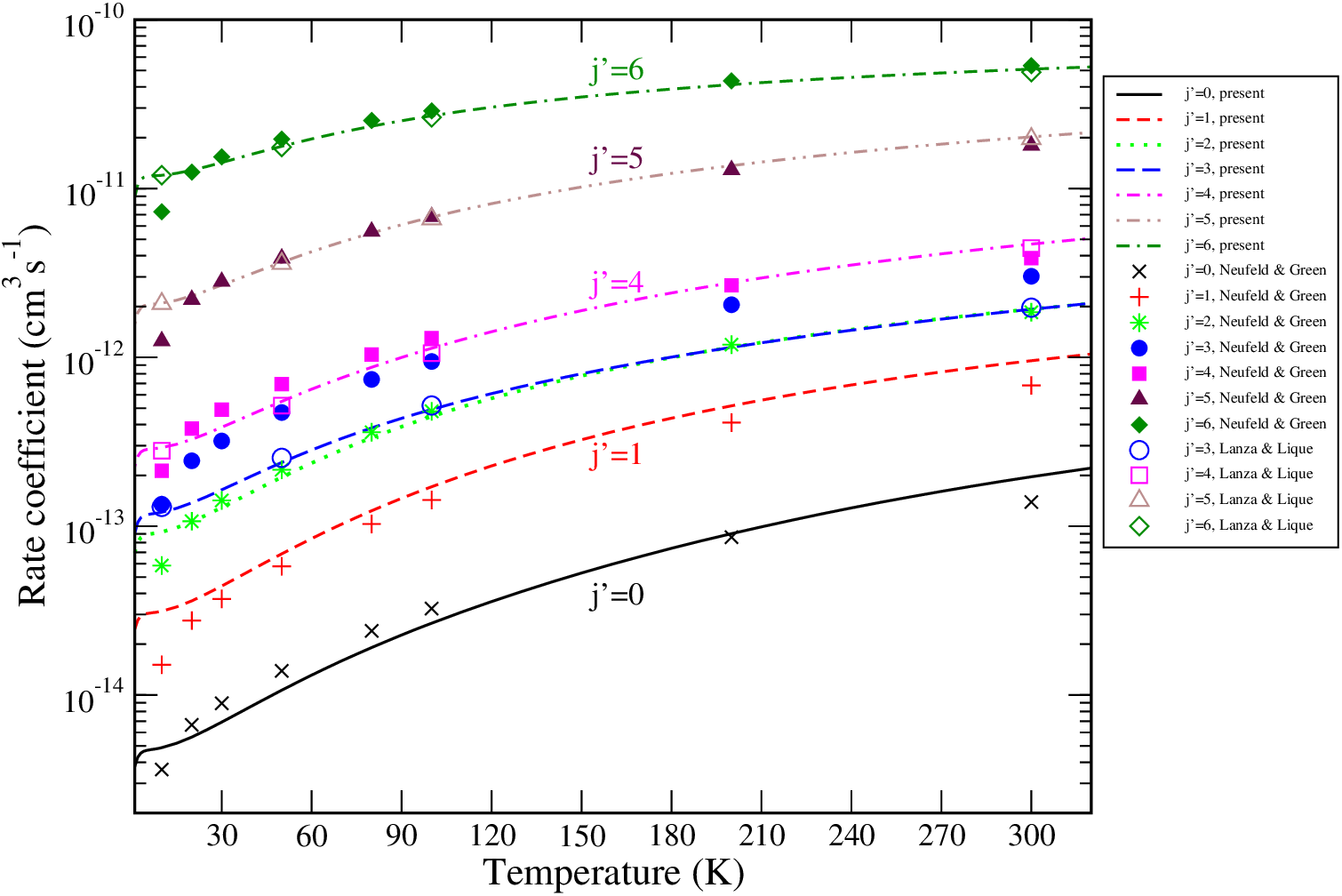}
\caption{ Same as Figure~\ref{fig3}, but for initial state $j=7$.}

\label{fig4}
\end{figure}

The state-to-state thermal rate coefficients can be calculated by  
thermally averaging the appropriate state-to-state cross sections over 
Maxwell-Boltzmann distribution of kinetic energy $E_j$. 
To our knowledge, there has been no published experimental rate coefficients available 
for rotational transitions of HCl by collisions with He. Therefore, we compare
our rate coefficients with the theoretical results of \citet{neu94}, \citet{lan12},
and \citet{aji13}, which were obtained over a limited temperature range of 10 to 300~K.
As examples, Figs.~\ref{fig3} and \ref{fig4} show selected transitions for 
state-to-state deexcitation rate coefficients from 
initial HCl levels $j$=1, 3, and 7.  Generally, our results 
show very good agreement with that of \citet{lan12} and \citet{aji13}, which were computed 
using different potentials.

For initial state $j=1$, Fig.~\ref{fig3}(a) shows that the present rate coefficients are larger than
those of \citet{neu94}, \citet{lan12}, and \citet{aji13} in the temperature range from 10 to 300 K,  but 
their rate coefficients approach our results with increasing temperature.
Fig.~\ref{fig3}(b), which displays state-to-state deexcitation rate coefficients  
 from initial level $j=3$,
 shows that the results of \citet{neu94}, \citet{lan12}, and \citet{aji13} agree well with the present
rate coefficients, except that for  quenching to $j'$=0 
the rate coefficients of both \citet{neu94} and \citet{lan12}  are larger than the
present results. 

State-to-state rate coefficients from  initial state $j=7$ 
are given in Fig.~\ref{fig4} where comparison of the present results with
the rate coefficients of \citet{lan12} show excellent agreement. Except for the
transition $j=5 \rightarrow j'=3$, the rate coefficients of \citet{neu94} generally
show good agreement with present results. 
These differences are likely related to differences in the adopted PESs (see below).
For illustration, in Fig.~\ref{fig5} we also present the state-to-state deexcitation rate coefficients
for temperatures from 0.1 K to 3000 K for initial levels $j$=10 and 20. 
Over the whole temperatures range considered, the rate coefficients generally
increase with increasing temperature for all transitions.
Further, one can clearly see that the rate coefficients decrease with increasing 
$|\Delta j|=|j'-j|$ with the $\Delta j=-1$ transitions dominant.

\section{Uncertainty in Scattering Calculations }

One essential prerequisite for accurate cross section and rate coefficient calculations is
the availability of a PES with high accuracy. 
However the quality of the PES depends not
only on the methods used for the interaction energy calculation, but also the accuracy and
reliability of the potential fitting and extension to long range. 
Convergence testing was performed in our calculations 
by using sufficiently large basis sets and adjustment of  
other calculation parameters. Consequently, the 
uncertainty associated with our cross section calculation 
is related primarily to the uncertainty in the adapted PES.  
In this work, the SAPT PES of \citet{mur04} was applied. 
As discussed by \citet{mur04}  the main features of the HCl-He
PES are a minimum in the linear configuration He-H-Cl, global minimum at the He-Cl-H
configuration, and a saddle point close to the T-shaped configuration. 
This also applies to PESs of \citet{faj06} and \citet{lan12}. 
However, the major difference between the three potentials is the well depth of the
global minimum. To estimate the uncertainty of the cross sections 
and rate coefficients calculated
from the PES of \citet{mur04}, a simple approach is to scale the SAPT PES 
by the largest and smallest ratio of the well depths of the three PESs. 
 Of course, the cross section and rate coefficients 
depend on the detailed structure of the potential including the shape,  well depth, 
and position of
the minumum which relies heavily on the methods applied in the HCl-He potential calculaltion.
Here, the well depth is used to illutrate the uncertainty in scattering calculations.
We calculated the deexcitation cross section and
rate coefficient for the $j=1 \rightarrow j'=0$ transition using the SAPT PES scaled 
by factors of 1.0322 and 0.9688, respectively, i.e., a variation of $\pm 3\%$.

In Fig.~\ref{fig6} we compare the $j=1 \rightarrow j'=0$
quenching cross sections and rate coefficients obtained using the scaled SAPT potential with
the results obtained from the original SAPT potential.
 Due to differences in the PES well-depths, it can be
seen that the cross sections and rate coefficients show evident disagreement at low collision
energy and low temperature, $T\le 20$ K. In the van der Waals interaction-dominated regime, the 
rate coefficients exhibit an oscillatory temperature dependence due to the
presence of resonances. 
The magnitudes, widths, and positions of the resonances are sensitive to the details of 
the PES.
However for high collision energy and high temperature, the cross
sections and rate coefficients from the scaled and original SAPT PES generally converge.
In another words, the uncertainty due to the accuracy of the SAPT PES is negligible at high
temperatures. 
We estimate the uncertainty in the computed rate coefficients to be 
12\%, 4\%, and  1.3\% at 1, 10, and 100 K, respectively.

\section{Astrophysical Applications}

Rate coefficients for collisional excitation and deexcitation are of importance in
describing the dynamics of energy transfer processes in interstellar objects. 
In particular, accurate rotational and vibrational excitation rates are needed 
to interpret microwave and infrared observations of the interstellar gas 
for non-LTE line formation.  Further, the thermal balance of interstellar gas 
is partly determined by cooling processes involving molecular
collisional excitation followed by radiative decay.
Despite some progress in laboratory measurements of state-to-state 
collisional rate coefficients and cross sections, astrophysical models depend almost
exclusively on theoretical data. Experiments do provide, when available, some
confidence in the theoretical rate coefficients.

As discussed in the Introduction, HCl has been observed in emission and absorption
in a variety of astronomical environments. 
Ag\'{u}ndez et al. (2011) reported observations of the $j$=1-0, 2-1, and 3-2
rotational lines of H$^{35}$Cl and H$^{37}$Cl in the carbon star envelope IRC+10216
using the {\it Herschel}/HIFI instrument. 
It was inferred that HCl is produced in the inner
layers of the envelope close to the asymptotic giant branch (AGB) star.
Recently, by observing the H$^{37}$Cl and  H$^{35}$Cl 1-0 transitions with the HIFI spectrometer, 
Codella et al. (2012) presented the first detection of HCl
towards protostellar shocks. 
It is expected that more highly excited rotational lines may be observed with SOFIA in the future.
However, except for the set of calculations done for  temperatures 
between 10 and 300 K and $j<7$  by \citet{neu94}, \citet{lan12}, and \citet{aji13},
no other computations of the He excitation rate coefficients have appeared.
Therefore, the current rate coefficient calculations,
which extend from 0.1 to 3000~K, are the most comprehensive
to date for He and can be utilized in a variety of applications augmenting the
datasets developed for He, H$_2$ and electrons \citep[see also, the LAMDA website,][]{sch05}.

 The rate coefficients for molecular scattering with para-H$_2$
 are often estimated using available rate coefficients with He by the
application of a constant
reduced-mass scaling factor of 1.4 \citep{sch05}. However, 
\citet{liq08} calculated the rate coefficients for SiS scattering with para-H$_2$ and compared
their results to the SiS-He rate coefficients \citep{vin07} 
 scaled by the H$_2$/He reduced-mass factor and found significant differences.
In much earlier work,
Schaefer (1990) found large deviations at temperatures of 100 K and below
when comparing directly calculated rotational excitation rate coefficients
for HD due to H$_2$ and those obtained by reduced-mass-scaling of
HD-He rates. Applying such a procedure to the current HCl-He rates coefficients
to estimate rate coefficients for H$_2$ is therefore not recommended. 

The hyperfine structure splitting of HCl occurs due to 
 its non-zero nuclear spin. Though
hyperfine transitions are not considered here, the hyperfine excitation cross sections can be
 estimated from the current results using an infinite-order sudden approximation approach \citep{lan12,fau12}.

\section{CONCLUSION}

Cross sections and rate coefficients for rotational quenching of HCl 
due to He collisions have been studied using the close-coupling method and the coupled-states 
approximation on the PES of \citet{mur04}  
for excited initial rotational levels of HCl up to $j$=20. 
State-to-state rate coefficients  
are obtained over a wide temperature range from 0.1 to 3000~K 
and available in tables formatted for astrophysical applications.  
The very good agreement with the results of 
\citet{lan12} and \citet{aji13}, computed on different PESs 
confirms the accuracy of the present calculations of the rate coefficients. 
The uncertainty in the current rate coefficient calculation,
deduced by scaling the adopted potential, is less than 4\% at temperatures of astrophysical interest,
comparable to the divergence in the well depth of recently available PESs.

\acknowledgements
This work was supported by NASA under Grant No. NNX12AF42G.

{}


\begin{table}
\begin{center}
\caption{Rotational excitation energies (cm$^{-1}$) of HCl for the ground 
  vibrational state\tablenotemark{a}
}
\vskip 0.5cm
\begin{tabular}{c c @{\hspace{1.0cm}} c c @{\hspace{1.0cm}}  c c }
\tableline \tableline
    {$j$} & {$E_j$} &  {$j$} & {$E_j$} & {$j$} & {$E_j$}    \\ [0.5ex]
\tableline
    0   &         0.00000   &   14   &      2201.22000   &   28   &      8252.30728   \\   [1ex]
    1   &        21.18448   &   15   &      2511.86400   &   29   &      8815.01400   \\   [1ex]
    2   &        63.54072   &   16   &      2842.16608   &   30   &      9393.37200   \\   [1ex]
    3   &       127.04328   &   17   &      3191.92272   &   31   &      9986.99968   \\   [1ex]
    4   &       211.65400   &   18   &      3560.91768   &   32   &     10595.50272   \\   [1ex]
    5   &       317.32200   &   19   &      3948.92200   &   33   &     11218.47408   \\   [1ex]
    6   &       443.98368   &   20   &      4355.69400   &   34   &     11855.49400   \\   [1ex]
    7   &       591.56272   &   21   &      4780.97928   &   35   &     12506.13000   \\   [1ex]
    8   &       759.97008   &   22   &      5224.51072   &   36   &     13169.93688   \\   [1ex]
    9   &       949.10400   &   23   &      5686.00848   &   37   &     13846.45672   \\   [1ex]
   10   &      1158.85000   &   24   &      6165.18000   &   38   &     14535.21888   \\   [1ex]
   11   &      1389.08088   &   25   &      6661.72000   &   39   &     15235.74000   \\   [1ex]
   12   &      1639.65672   &   26   &      7175.31048   &   40   &     15947.52400   \\   [1ex]
   13   &      1910.42488   &   27   &      7705.62072   &        &                   \\   [1ex]
\tableline
\end{tabular}
  \tablenotetext{a}{$E_j = B_0j(j+1)-D_0j^2(j+1)^2$, where $B_0$ and $D_0$ are the rotational constant  
  and the centrifugal distortion constant of HCl, respectively.}
\end{center}
\label{tab1}
\end{table}

\begin{figure}
\epsscale{0.9}
\plotone{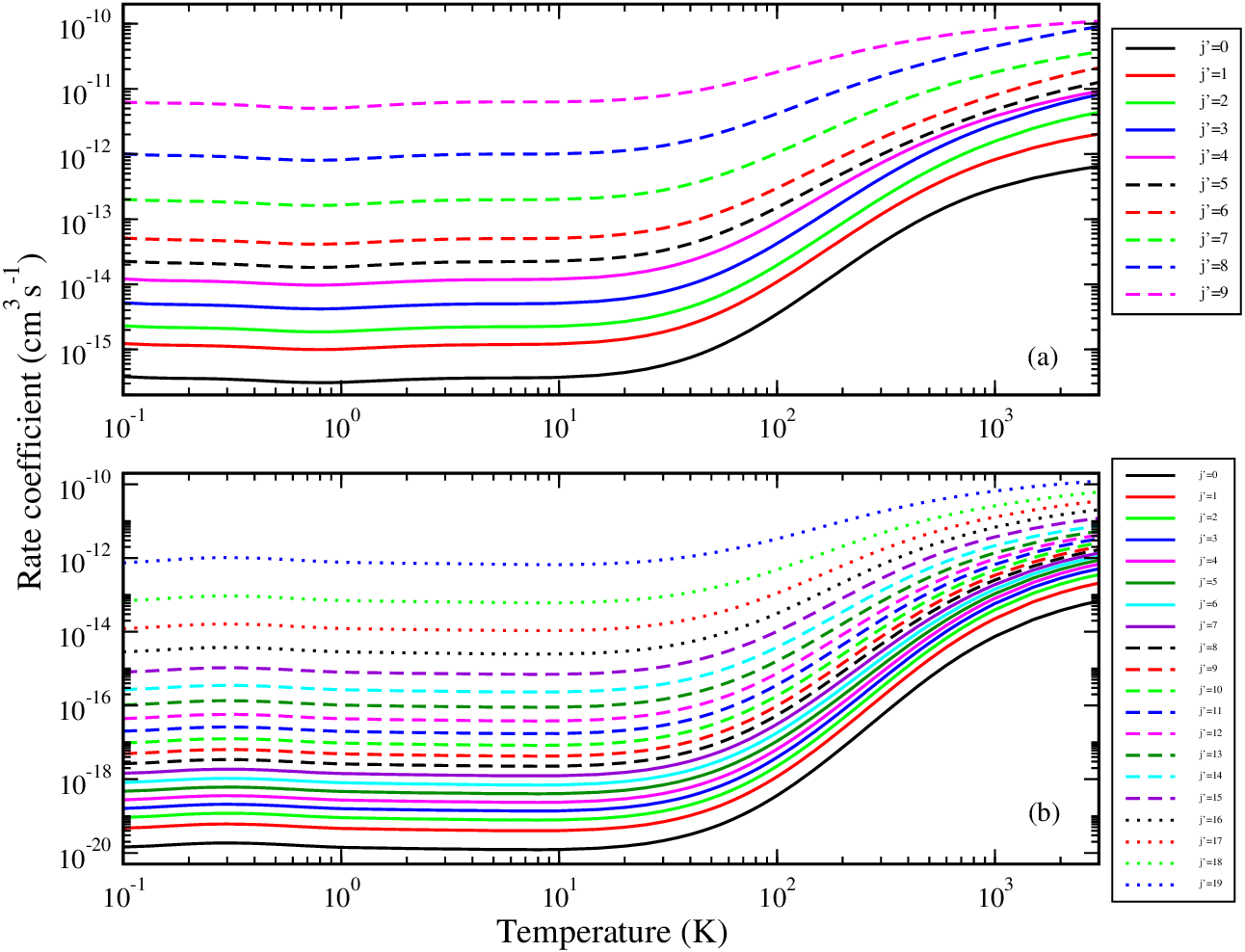}
\caption{ 
State-to-state rotational deexcitation rate coefficients from initial state $j=10$ 
and $j=20$ of HCl due to collisions with He atoms as a function of temperature.  (a) $j=10$, (b) $j=20$.
}
\label{fig5}
\end{figure}

\begin{figure}
\epsscale{0.9}
\plotone{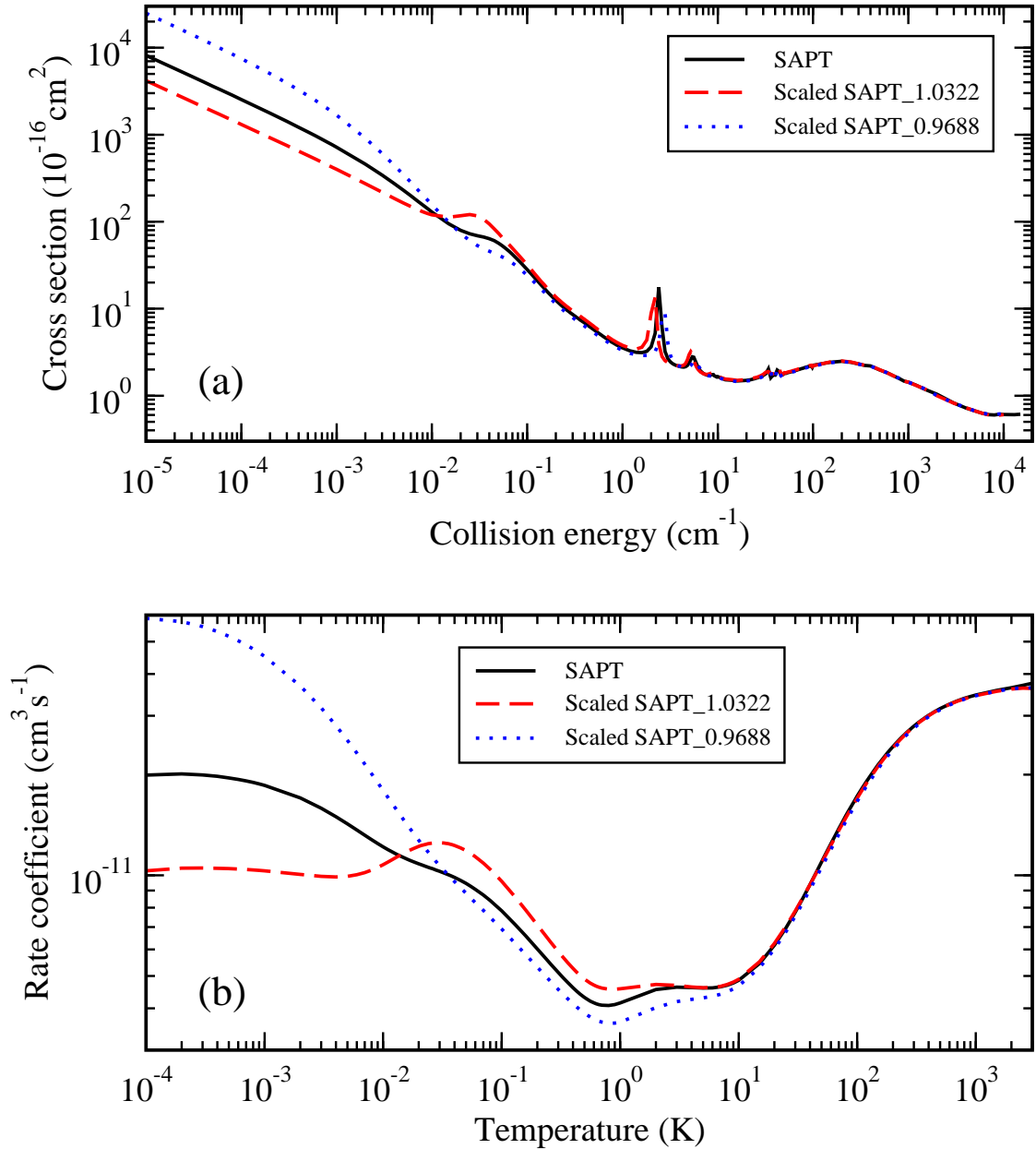}
\caption{ 
Rotational deexcitation cross section (a) and rate coefficient (b) for 
$j=1 \rightarrow j'=0$ 
using the SAPT potential and scaled SAPT potentials.
Solid line: SAPT potential of \citet{mur04}, dashed line: scaled SAPT potential with factor of 1.0322, 
dotted line: scaled SAPT potential with factor of 0.9688.
}
\label{fig6}
\end{figure}

\end{document}